\documentclass[9pt,conference]{IEEEtran}

\usepackage[preprint]{waspaa25}

\usepackage{amsmath,amsfonts,bm}
\usepackage{algorithmic}
\usepackage{array}
\usepackage[caption=false,font=normalsize,labelfont=sf,textfont=sf]{subfig}
\usepackage{textcomp}
\usepackage{stfloats}
\usepackage{url}
\usepackage{verbatim}
\usepackage{graphicx}
\usepackage{booktabs,tabularx}
\usepackage{balance}
\usepackage{hyperref}
\usepackage{svg}

\begin{document}
\title{Miipher-2: A Universal Speech Restoration Model \\ for Million-Hour Scale Data Restoration}

\name{Shigeki Karita, Yuma Koizumi, Heiga Zen, Haruko Ishikawa, Robin Scheibler, Michiel Bacchiani}
\address{Google DeepMind, Tokyo, Japan}

\maketitle

\begin{abstract}
Training data cleaning is a new application for generative model-based speech restoration (SR). This paper introduces Miipher-2, an SR model designed for million-hour scale data, for training data cleaning for large-scale generative models like large language models. Key challenges addressed include generalization to unknown languages, operation without explicit conditioning (e.g., text, speaker ID), and computational efficiency. Miipher-2 utilizes a frozen, pre-trained Universal Speech Model (USM), supporting over 300 languages, as a robust, conditioning-free feature extractor. To optimize efficiency and minimize memory, Miipher-2 incorporates parallel adapters for predicting clean USM features from noisy inputs and employs the WaveFit neural vocoder for waveform synthesis. These components were trained on 3,000 hours of multi-lingual, studio-quality recordings with augmented degradations, while USM parameters remained fixed. Experimental results demonstrate Miipher-2's superior or comparable performance to conventional SR models in word-error-rate, speaker similarity, and both objective and subjective sound quality scores across all tested languages. Miipher-2 operates efficiently on consumer-grade accelerators, achieving a real-time factor of 0.0078, enabling the processing of a million-hour speech dataset in approximately three days using only 100 such accelerators.
\end{abstract}


\section{Introduction}
\label{sec:introduction}

Speech restoration (SR) refers to the process of transforming degraded speech signals into their high-fidelity counterparts~\cite{Maiti_waspaa_2019,self_remaster,Su_2021,liu22y_interspeech,UNIVERSE,koizumi_2023_sr,Richter_2023,diffusion_sr_review_2024,scheibler24_interspeech,kang2025,genhancer,liu2024jointsemanticknowledgedistillation,ditse}. Recently, the application of generative models to SR tasks has become increasingly prevalent. These advancements enable SR methodologies to effectively mitigate diverse acoustic degradations, such as noise, reverberation, and codec artifacts, producing high-quality audio comparable to professional studio recordings~\cite{Maiti_waspaa_2019,self_remaster,Su_2021,liu22y_interspeech,UNIVERSE,koizumi_2023_sr,Richter_2023,diffusion_sr_review_2024,scheibler24_interspeech,kang2025,genhancer,liu2024jointsemanticknowledgedistillation,ditse}.

This progress has facilitated a novel application domain for SR: data cleaning for Text-to-Speech (TTS) training datasets. Koizumi \textit{et al.} proposed Miipher~\cite{koizumi_2023_sr}, a monolingual robust SR model for English conditioned on textual and speaker identity information. Their research demonstrated the feasibility of restoring potentially noisy public datasets to studio-level quality, thereby enabling the training of high-performance TTS models using these enhanced corpora~\cite{librittsr_koizumi23_interspeech,fleursr}.

The performance of generative models, including Large Language Models (LLMs)~\cite{gemini,openai2024gpt4technicalreport,moshi}, is critically dependent on the volume and quality of the training data, underscoring the importance of research into data quality enhancement. Large-scale training datasets are frequently acquired via web-scraping, a process inherently prone to introducing noisy samples. Consequently, for text and image modalities, quality filtering techniques are commonly employed to curate cleaner datasets~\cite{gemini,openai2024gpt4technicalreport,moshi}. However, obtaining clean speech recordings from web sources is more difficult than other modalities due to the nature of sound, i.e. inherent contamination from interference sources and reverberation.

The application of SR for cleaning web-scraped and million-hour scale speech datasets introduces several novel challenges:
\begin{itemize}
    \item Handling unknown languages: For low-resource languages, sufficient studio-quality speech data for training SR models may be unavailable. Thus, the SR model must be capable of processing languages for which dedicated high-quality training data is absent.
    \item Conditioning free inference: Manual annotation of transcription and/or speaker ID for large-scale datasets is often infeasible or cost-prohibitive. Therefore, the SR model must operate directly on the waveform without reliance on external conditioning features.
    \item Computational efficiency: Inference latency must be small while preserving output quality. Furthermore, achieving affordable large-scale parallel processing necessitates minimizing the memory footprint on hardware accelerators.
\end{itemize}

We propose Miipher-2, a universal SR model designed for enhancing giant-scale speech datasets. To enable operation in languages lacking dedicated SR training data, we employ the Universal Speech Model (USM), a self-supervised learning (SSL) model pre-trained on noisy data spanning over 300 languages by a prior work~\cite{usm}, as a frozen feature extractor. We observe that leveraging an SSL model trained on such extensive data obviates the need for the text and speaker conditioning previously required~\cite{koizumi_2023_sr}. For computational efficiency, we replace the auxiliary feature cleaner network utilized in~\cite{koizumi_2023_sr} with parallel adapters (PAs)~\cite{paralleladapter}. Additionally, to reduce memory consumption, we introduce modifications to the WaveFit~\cite{wavefit} vocoder architecture. PAs and WaveFit were trained on about 3,000 hour, 54-language dataset of studio-quality recordings with added artificial noise and reverberation. Our experiments demonstrate that Miipher-2 achieves a fast real-time factor (RTF) of 0.0078 on a smallest TPU v4i chip with 8 GB device memory~\cite{tpuv4i}, enabling the processing of a million-hour speech dataset in approximately 3 days using only 100 TPU v4i chips in-parallel. Furthermore, Miipher-2 performs comparably to the original Miipher model~\cite{koizumi_2023_sr} on English data and achieves similar quality scores for unknown languages.
Audio samples of the restored samples are available at our demo page
\footnote{\url{https://google.github.io/df-conformer/miipher2}}.

\section{Universal Speech Restoration Model}
\label{sec:proposed_method}

\begin{figure}
    \centering
        \includegraphics[width=0.48\textwidth]{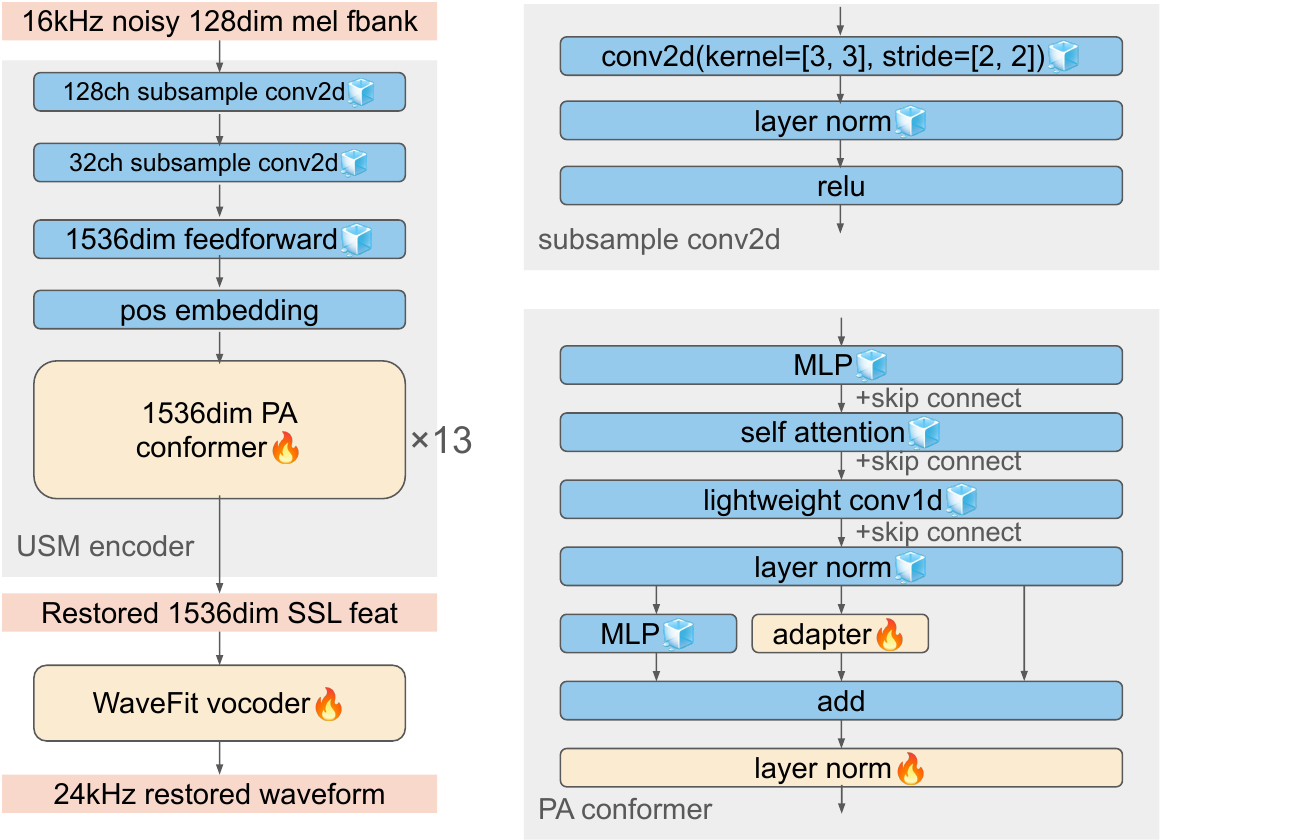}
    \vspace{-10pt}
    \caption{Miipher-2 architecture overview and its USM encoder details. Blue blocks are frozen or non-learnable, while a few orange blocks in conformer blocks and the WaveFit vocoder are fintuned.}
    \label{fig:miipher2}
\end{figure}

Miipher-2 is a generative SR based on based on parametric resynthesis strategy~\cite{Maiti_waspaa_2019}. An overview of Miipher-2 is shown in Fig.~\ref{fig:miipher2}, which comprising two primary components: a feature cleaner, which predicts acoustic features corresponding to a clean waveform from an input noisy waveform, and a vocoder, which subsequently synthesizes a waveform from these predicted clean features.

\subsection{Feature extractor model}

Utilizing SSL models for feature extraction in SR is known to be effective~\cite{koizumi_2023_sr,kang2025,genhancer,liu2024jointsemanticknowledgedistillation,ditse}. Prior researches indicate that an SSL model trained on diversely degraded data can yield an SR model robust to unknown sound quality issues, even when other SR components are trained on limited simulated noisy data~\cite{koizumi_2023_sr}.

We hypothesized this framework's applicability extends beyond degradation patterns to encompass unknown languages. While acquiring studio-quality recordings for all languages is impractical, collecting noisy multilingual speech is feasible. Consequently, we employ the USM, an SSL model pre-trained on 12 million hours of YouTube audio spanning over 300 languages.

We also posit that the BEST-RQ~\cite{bestrq} style fixed-random quantizer within USM is effective for eliminating the need for textual or speaker-ID conditioning. Contrastive loss for training codebooks in wav2vec 2.0~\cite{w2v22020} and w2v-BERT~\cite{w2vbert} result in each codebook representing specific phonemes. Although this may improve speech recognition accuracy, it potentially discards crucial features like speaking style variations, and low-frequency phonemes of low-resource languages by focusing on typical phonemes.
In addition, its negative sampling minimizes similarity between masked and non-masked units in the same utterance, which can make the units insensitive to speaker and acoustic environments.
Conversely, BEST-RQ masked token prediction learning with its frozen random quantizer, is expected to retain finer-grained acoustic information. This may be the reason why HuBERT~\cite{hubert} and WavLM~\cite{Chen2021WavLM}, which do not have trainable codebook nor contrastive loss, are successful in SR tasks~\cite{ditse,baas2023knnvc}.

Based on these assumptions, we use a non-fine-tuned 2-billion parameter version of the USM~\cite{usm}. The 13th layer was selected for intermediate feature extraction, guided by preliminary experiments and the observation that deeper layers in SSL speech feature extraction tend to lose fine-grained acoustic information~\cite{zhu2023vectokspeechspeechvectorization,baas2023knnvc}.

\subsection{Parameter-efficient feature cleaner}

To predict USM features corresponding to clean waveforms, parallel adapters (PA) are employed~\cite{paralleladapter}\footnote{The reason why we adopt PA rather than LoRA etc was strong performances in USM downstream tasks e.g. speech recognition, translation~\cite{usm}.}. These adapters consist of feed-forward network (FFN) layers appended to each USM layer. The raw USM output is summed with the adapter output, serving as the input to the subsequent layer. The utilization of PAs, rather than the DF-Conformer~\cite{Koizumi_waspaa_2021} implemented in Miipher~\cite{koizumi_2023_sr}, aims to reduce the number of trainable parameters and improve inference speed—a critical factor for processing large-scale datasets. While the feature cleaner in Miipher contained 100M parameters, the PAs in Miipher-2 comprise only 20M parameters. Furthermore, the absence of attention layers in the adapters results in linear computational complexity with respect to sequence length, facilitating faster inference.

\begin{figure}
    \centering
        \includegraphics[width=0.4\textwidth]{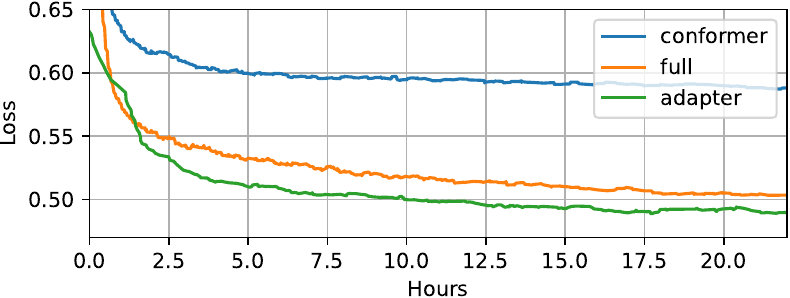}
    \vspace{-8pt}
    \caption{Loss curves over time for three feature cleaner training strategies: Miipher-1 conformer based cleaner (conformer),  updating all USM parameters (full), and porposed PA fine-tuning (adapter).%
    }
    \label{fig:sftloss}
\end{figure}

The loss function is the same as Miipher~\cite{koizumi_2023_sr}, a sum of L1, L2 and spectral convergence~\cite{spectralconvergence} loss values between predicted and target clean 13-th USM layer features. Figure~\ref{fig:sftloss} illustrates the loss curves over time (hours) for several approaches: the Miipher's Conformer cleaner, updating all USM parameters, and 1024-dimension PA fine-tuning. It was observed that PA converges most rapidly because PA requires updating fewer trainable parameters (only 3\%) relative to full USM retraining, and the lowest absolute loss value was smaller than full USM finetuning.

\subsection{Memory-efficient WaveFit}

\begin{figure}
    \centering
    \includegraphics[width=0.35\textwidth]{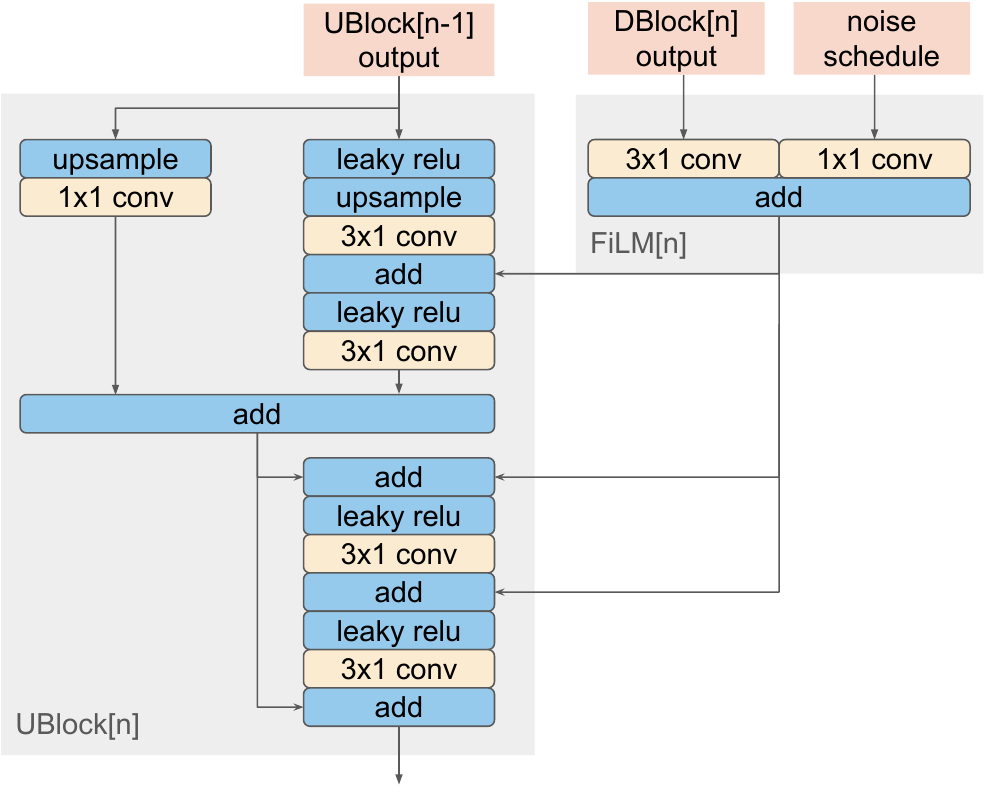}
    \vspace{-8pt}
    \caption{Memory efficient FiLM and UBlock in Miipher-2 WaveFit.}
    \label{fig:unet}
\end{figure}

We also implement two improvements for memory efficiency for WaveFit~\cite{wavefit}. The first improvement is a feature pre-upsampler. Miipher utilized transposed convolution-based upsampling because the w2v-BERT~\cite{w2vbert} feature frame rate (25 Hz) differs from the mel-spectrogram frame rate (80 Hz). This method led to substantial memory consumption, mainly from extra padding required by TPUs, as the channel dimension had to be a multiple of 128. To eliminate this restriction, transposed convolutions were replaced with an upsampling module that repeats the USM output four times along the time axis.

The second improvement targets the FiLM (Feature-wise Linear Modulation) layer~\cite{film}, the second most memory-intensive component. Its high memory usage stems from generating an output dimension six times larger than its input (two scaling and biasing parameters for three feature-wise Affine operations in UBlock~\cite{wavegrad}) and retaining its longest output sequence throughout the U-Net. Therefore, the first FiLM layer, connecting the initial DBlock to the final UBlock and processing the longest input, was removed.
Further memory optimization was achieved by simplifying the UBlock: the shared FiLM output is directly added to the hidden vectors, instead of applying three distinct feature-wise Affine operations, as shown in Fig.~\ref{fig:unet}.

\section{Experiments}
\label{sec:eval}

\subsection{Comparison methods}

We compared Miipher-2 with one public SR model and three Miipher variants.

\noindent\textbf{TF-GridNet:} The baseline model for the URGENT Challenge 2025~\cite{urgent2025}. Please noted that this serves only as reference of a publicly available SR model since training dataset is different.

\noindent\textbf{Miipher-1:} A monolingual text and speaker-feature conditioned SR model for English~\cite{koizumi_2023_sr}. The LibriTTS-R dataset~\cite{librittsr_koizumi23_interspeech} cleaned by this model demonstrated high performance in multiple TTS papers~\cite{prompttts,casanova24_interspeech,lyth2024naturallanguageguidancehighfidelity}. Therefore, results comparable or superior to this model would affirm its sufficient performance for data cleaning tasks.

\noindent\textbf{Miipher-USM:} A Miipher-2 model variant lacking computational efficiency improvements to evaluate potential SR performance degradation from our proposal. This variant employs the standard Conformer-based feature cleaner and vanilla WaveFit.

\noindent\textbf{Miipher-2-P:} The Miipher-2 model trained on public datasets cleaned by Miipher-2. We evaluate if such data can achieve performance equivalent to high-quality studio recordings. The cleaned multilingual datasets include CoVoST1~\cite{wang-etal-2020-covost}, CVSS~\cite{jia2022cvss}, Multilingual LibriSpeech (MLS)~\cite{mls_pratap20_interspeech}, and FLEURS~\cite{fleurs}.

\subsection{Experimental condition}

Miipher-2 and Miipher-USM were trained on simulated noisy-clean paired data. The clean data comprised 3,195 hours of speech from 1,642 speakers across 44 languages (54 locales). The noise dataset consisted of internally collected audio snippets from environments such as cafes, kitchens, and automobiles. Noisy utterances were synthesized by mixing randomly selected speech and noise samples, with signal-to-noise ratios (SNRs) ranging from 5 dB to 30 dB. This noisy dataset was augmented using four patterns, determined by the presence or absence of reverberation and codec artifacts, following~\cite{UNIVERSE}. A unique room impulse response (RIR) for each sample was generated via a stochastic RIR employing the image-source method~\cite{image_method}. Parameters for the stochastic RIR and codecs were consistent with~\cite{koizumi_2023_sr}.

PA was configured with a 1024 hidden dimension and 1532 input/output dimensions at each post-feedforward layer of USM. WaveFit converted USM output to waveforms, employing a pre-network of four 1532-dimension conformer layers (similar to USM) followed by a fixed-point iteration U-Net. The U-Net utilized 2/2/3/4 downsampling on 128/128/256/512 dimensions and 5/4/3/2/2 upsampling on 512/512/256/128/128 dimensions. Its GAN and STFT loss functions were identical to Miipher-1~\cite{koizumi_2023_sr}.

Initially, PA was trained for 800k steps. Subsequently, WaveFit was pre-trained for 200k steps to predict clean waveforms from USM features extracted from clean waveforms. Finally, the pre-trained WaveFit was fine-tuned for 675k steps to predict clean waveforms from clean features predicted by USM and PA, using noisy waveform inputs. The optimizer configuration followed~\cite{Su_2020}, with a batch size of 512.

\subsection{Computational efficiency}
\label{sec:eval:efficiency}

\begin{table}[tb]
    \caption{Peak device memory usage [MB] and real-time factor (RTF) on TPU v4i using Miipher-USM vs Miipher-2 using bfloat16 activations for 30sec 16kHz speech restoration. OOM denotes out-of-memory.}
    \label{tab:hbm}
    \centering
    \begin{tabular}{crrrr}
\toprule
          & \multicolumn{2}{c}{\textbf{Memory} $(\downarrow)$ } & \multicolumn{2}{c}{\textbf{Real-Time Factor} $(\downarrow)$}  \\
Batch & Miipher-USM & Miipher-2 & Miipher-USM & Miipher-2  \\
\midrule
   1 	& {5612.94}	& \textbf{2694.98} & 0.0565 &	\textbf{0.0555} \\
   2	& OOM	& \textbf{3228.50} &	OOM &	\textbf{0.0253} \\
   4	& OOM	& \textbf{4434.69} &	OOM &	\textbf{0.0130} \\
   8	& OOM	    & \textbf{6635.06}  & OOM & \textbf{0.0078} \\
\bottomrule
    \end{tabular}
\end{table}

We first evaluated the computational efficiency improvement by comparing Miipher-2 and Miipher-USM. Tables~\ref{tab:hbm} enumerates the peak memory usage and real-time factor (RTF) for inference of a 30-second, 16kHz speech segment on a smallest TPU (v4i) with 8GB of device memory~\cite{tpuv4i}. Miipher-USM's inefficient memory utilization limits processing to a single sample, yielding an inference batch size of 1. Meanwhile, Miipher-2 reduces memory usage by 52\% with a batch size of 1, allowing its inference batch size to increase to 8. Consequently, Miipher-2 processes 240 seconds of data ($30$ sec $\times$ $8$ sample) per inference, achieving an RTF of 0.0078. This is a 724\% speed improvement over Miipher-USM. This RTF enables cleaning one million hours of data in approximately 3 days using 100 consumer-grade TPUs, demonstrating the proposed computational efficiencies make million-hour-scale dataset cleaning feasible.

\subsection{Objective evaluation}
\label{sec:eval:sound_quality}

\begin{table}[ttt]
\caption{LibriTTS speech restoration automatic evaluation.}
\label{tab:eval:sound_quality:libritts}
\centering
\scriptsize
\begin{tabular}{l | c c c c}
\toprule
    & \textbf{DNSMOS} $(\uparrow)$ & \textbf{SQuId} $(\uparrow)$ & \textbf{WER} $(\downarrow)$ & \textbf{SPK} $(\uparrow)$ \\
\midrule
LibriTTS & 2.68 $\pm$ 0.010 & 3.85 $\pm$ 0.006 & 0.132 & N/A \\
TF-GridNet~\cite{urgent2025} & 2.67 $\pm$ 0.010 & 3.82 $\pm$ 0.006 & 0.136 & 0.945 \\
\midrule
Miipher-1~\cite{librittsr_koizumi23_interspeech} & 2.71 $\pm$ 0.010 & \textbf{4.02 $\pm$ 0.005} & 0.150 & 0.585 \\
Miipher-USM & 2.85 $\pm$ 0.009 & 4.01 $\pm$ 0.005 & 0.150 & 0.722 \\
Miipher-2 & \textbf{2.87 $\pm$ 0.009} & 4.00 $\pm$ 0.005 & \textbf{0.149} & 0.744 \\
Miipher-2-P & 2.79 $\pm$ 0.010 & 3.95 $\pm$ 0.006 & 0.154 & \textbf{0.746} \\
\bottomrule
\end{tabular}
\end{table}

We evaluated overall speech restoration performance in four automated evaluation metrics: word-error-rate (WER) by a single multilingual ASR model using USM encoder finetuned with CTC decoder~\cite{ctc} and language ID embedding, speaker similarity (SPK) using dthe speaker embedding~\cite{NEURIPS2018_6832a7b2,chen2018sample}, and two types of predicted mean-opinion-score (MOS) using DNSMOS~\cite{reddy2021dnsmos} and SQuId~\cite{squid}. To evaluate the SR performance on training data for actual speech generative models, 500 samples randomly selected from the LibriTTS test-other dataset were used for evaluation.
Note that scores of Miipher-1 were calculated from LibriTTS-R dataset~\cite{librittsr_koizumi23_interspeech}.

Table~\ref{tab:eval:sound_quality:libritts} shows the objective evaluation results. Miipher-2 achieved comparable performance to the text-conditioned, English-only Miipher-1 model in predicted MOS and WER. Furthermore, Miipher-2-P, trained on a Miipher-2-processed public multilingual dataset, attained nearly equivalent performance, indicating Miipher-2's efficacy for multilingual dataset cleaning. Compared to Miipher-USM, Miipher-2 demonstrated statistically significant DNSMOS/SPK improvements ($p<0.001$, $t$-test) with no statistically significant SQuId/WER degradation ($p>0.5$), suggesting the proposed computational efficiency enhancements do not compromise performance. For SPK, Miipher-2 and its variants achieved significantly higher scores than Miipher-1 ($p=0$). This is likely attributable to USM enabling the use of acoustic features containing fine-grained details.

TF-GridNet~\cite{tfgridnet} baseline trained on WSJ~\cite{wsj} and Common Voice~\cite{common_voice} corpora by URGENT2025 challenge~\cite{urgent2025}, which preserves WER and SPK as original data but it could not improve DNSMOS and SQuId at all. A potential explanation is that this model was trained on a public dataset wherein the target data may have possessed insufficient cleanliness. Given the high performance of Miipher-2-P, employing datasets cleaned by an SR model for training subsequent SR models should be well worth further investigation.

\subsection{Subjective evaluation}

Subjective quality was evaluated using  mean-opinion-score (MOS) and side-by-side (SxS) same as Miipher-1 based LibriTTS-R report~\cite{librittsr_koizumi23_interspeech}. The scale of human MOS was a 5-point scale (1: Bad, 2: Poor, 3: Fair, 4: Good, 5: Excellent) with rating increments of 0.5. In SxS evaluation comparing Miipher-2 with other methods, we simply asked ``Which sample has better quality?'' with a 7-point scale (-3: Much worse, -2: Worse, -1: Slightly worse, 0: About the same, 1: Slightly better, 2: Better, 3: Much better) with increments of 0.5 and random left/right flipping. Each subject was allowed to evaluate up to six stimuli, that is, 388 human reviewers participated in this experiment to evaluate 500 samples in LibriTTS test-other set.

We excluded TF-GridNet because it did not yield a statistically significant quality improvement from the original noisy samples ($p>0.2$, $t$-test). To reduce subject burden, Miipher-USM and Miipher-P were also omitted, as their performance was comparable to or marginally inferior to Miipher-2, leading to an expectation that their scores would lie between those of Miipher-1 and Miipher-2.

Table~\ref{tab:humaneval:sound_quality:libritts} shows the results of subjective evaluation.
Our Miipher-2 showed significantly better MOS than Miipher-1 ($p=0.008$) but SxS results in the subtle improvement ($p=0.433$).

\begin{table}[ttt]
\caption{LibriTTS restoration human evaluation with 95\% confidence interval. A positive SxS score indicates that Miipher-2 was preferred.}
\label{tab:humaneval:sound_quality:libritts}
\centering
\begin{tabular}{l | c c}
\toprule
\textbf{Method}& \textbf{MOS} $(\uparrow)$ & \textbf{SxS} \\
\midrule
 LibriTTS        & $2.81 \pm  0.118$ & $1.208 \pm 0.1280$ \\ %
\midrule
 Miipher-1~\cite{librittsr_koizumi23_interspeech} & $3.26 \pm 0.112$ & $0.044 \pm 0.1100$ \\ %
 Miipher-2       &  \textbf{3.46 $\pm$ 0.106} & N/A \\
\bottomrule
\end{tabular}
\end{table}

\subsection{Multilingual SR evaluation}

\subsubsection{Objective evaluation}

To evaluate multilingual capability, SR was conducted on both known and unknown languages. The MLS and FLEURS datasets served as test sets for known and unknown languages, respectively. Results are presented in Tables~\ref{tab:eval:sound_quality:known} for known languages and Table~\ref{tab:eval:sound_quality:unknown} for unknown languages. For both language categories, results were largely consistent with the English results shown in Table~\ref{tab:eval:sound_quality:libritts}. Specifically, predicted MOSs improved, WER slightly decreased, and SPK was approximately 0.7. These findings indicate Miipher-2's effective SR performance on both known and unknown languages.

The elevated WER for some unknown languages is attributed to the low performance of the ASR model itself on these low-resource languages. Given that Miipher-2 minimally impacts WER, it is inferred that Miipher-2 can effectively perform SR for these languages. languages as well.
One plausible reason of the lowest speaker similarity in pt\_pt is contamination of multi-speaker cases in its test set.

\begin{table}[ttt]
\caption{Speech restoration results of known locales from MLS test sets. (Original $\to$ Miipher-2)}
\label{tab:eval:sound_quality:known}
\centering
\scriptsize
\begin{tabular}{l | c c c c}
\toprule
& \textbf{DNSMOS} $(\uparrow)$ & \textbf{SQuId} $(\uparrow)$ & \textbf{WER} $(\downarrow)$ & \textbf{SPK} $(\uparrow)$ \\
\midrule
 de\_de  & $2.96 \to \bm{3.09} $ & $3.77 \to \bm{3.84}$ &  $\bm{9.23} \to 10.2$ & $0.709$ \\
 nl\_nl  & $2.99 \to \bm{3.03} $ & $3.81 \to \bm{3.83}$ & $\bm{10.3} \to 10.9$ & $0.788$ \\
 fr\_fr & $3.05 \to \bm{3.17} $ & $3.74 \to \bm{3.79}$ & $\bm{15.6} \to 19.4$ & $0.714$ \\
 es\_es & $3.06 \to \bm{3.17} $& $3.83 \to \bm{3.92}$ &  $\bm{4.85} \to  5.10$ & $0.730$ \\
 it\_it & $3.02 \to \bm{3.18} $& $3.62 \to \bm{3.73}$ & $\bm{13.6} \to 14.1$ & $0.701$ \\
 pt\_pt & $2.93 \to \bm{3.12} $ & $3.83 \to \bm{4.03}$ &  $\bm{7.65} \to  8.53$ & ${0.609}$ \\
 pl\_pl & $3.03 \to \bm{3.16} $ & $3.98 \to \bm{4.05}$ &  $\bm{4.90} \to  5.74$ & $0.754$ \\
\bottomrule
\end{tabular}
\vspace{8pt}
\caption{Speech restoration results of unknown locales from FLEURS test sets. (Original $\to$ Miipher-2)}
\label{tab:eval:sound_quality:unknown}
\centering
\begin{tabular}{l | c c c c}
\toprule
& \textbf{DNSMOS} $(\uparrow)$ & \textbf{SQuId} $(\uparrow)$ & \textbf{WER} $(\downarrow)$ & \textbf{SPK} $(\uparrow)$ \\
\midrule
  ca\_es
 & $2.87 \to \bm{3.12}$ & $3.75 \to \bm{3.96}$ & $\bm{5.01} \to 5.46$ & 0.637 \\
  ru\_ru
 & $2.72 \to \bm{2.95}$ & $3.69 \to \bm{3.87}$ & $\bm{5.25} \to 5.52$ & 0.601 \\
 ur\_pk
 & $2.87 \to \bm{3.03}$ &  $4.01 \to \bm{4.20}$  &$\bm{21.0} \to 22.1$ & 0.732 \\
 sw\_ke
 & $2.61 \to \bm{2.94}$ & $ 3.57 \to \bm{3.77}$ & $\bm{33.5} \to 35.2$ & $0.738$ \\
 mi\_nz
 & $2.69 \to \bm{3.03}$ & $3.16 \to \bm{3.40}$ & $\bm{38.4} \to 40.7$ & 0.569 \\
\bottomrule
\end{tabular}
\end{table}

\subsubsection{Comparison with Miipher-2-P on multilingual dataset}
\label{sec:eval:genmodel:sr}

To demonstrate the comparability of Miipher-2 cleaned data to studio-recorded speech for training non-English generative models, we evaluated Miipher-2 and Miipher-2-P on an internal 52-language noisy-clean paired dataset (Table~\ref{tab:public-vs-v2}). SPK values are higher and more accurate as clean speech was used for similarity computation with the restored audio. Overall, Miipher-2 and Miipher-2-P exhibited similar performance, though Miipher-2 achieved superior speaker similarity and SQuId in most languages. Conversely, Miipher-2-P outperformed Miipher-2 on CER/WER in four languages (fr\_fr, hu\_hu, km\_kh, vi\_vn), potentially due to it's training data encompassing more languages (e.g., FLEURS contains over 100 languages). This result indicates that such distillation is also beneficial for self-supervised training on new, unknown datasets.

\begin{table}
\caption{Internal multilingual speech restoration evaluation using Miipher-2 (v2) and Miipher-2-P (pub).}
\scriptsize
\label{tab:public-vs-v2}
\begin{tabular}{l|cccccccc}
\toprule
& \multicolumn{2}{c}{\textbf{DNSMOS} $(\uparrow)$ }& \multicolumn{2}{c}{\textbf{SQuId} $(\uparrow)$} &  \multicolumn{2}{c}{\textbf{WER} $(\downarrow)$} & \multicolumn{2}{c}{\textbf{SPK} $(\uparrow)$} \\
 & v2 & pub & v2 & pub & v2 & pub & v2 & pub \\
\midrule
af\_za & \textbf{2.97} & 2.92 & \textbf{4.09} & 4.01 & \textbf{23.49} & 24.22 & \textbf{0.873} & 0.790 \\
am\_et & \textbf{3.08} & 3.03 & \textbf{4.22} & 4.06 & \textbf{31.08} & 34.33 & \textbf{0.937} & 0.845 \\
ar\_eg & \textbf{3.08} & 3.02 & \textbf{4.34} & 4.22 & \textbf{15.20} & 17.69 & \textbf{0.914} & 0.810 \\
ar\_xa & \textbf{3.09} & 3.05 & \textbf{4.34} & 4.21 & \textbf{16.15} & 19.36 & \textbf{0.926} & 0.818 \\
bn\_bd & \textbf{3.04} & 2.96 & \textbf{4.38} & 4.25 & \textbf{24.75} & 25.71 & \textbf{0.931} & 0.841 \\
bn\_in & \textbf{3.05} & 2.96 & \textbf{4.38} & 4.24 & \textbf{24.93} & 25.60 & \textbf{0.931} & 0.840 \\
cmn\_cn & \textbf{3.10} & 3.05 & \textbf{4.32} & 4.23 & \textbf{41.28} & 42.57 & \textbf{0.936} & 0.843 \\
cmn\_tw & \textbf{3.06} & 3.05 & \textbf{4.22} & 4.17 & \textbf{15.66} & 17.18 & \textbf{0.929} & 0.846 \\
da\_dk & \textbf{3.01} & 2.93 & \textbf{4.18} & 4.09 & \textbf{22.13} & 24.03 & \textbf{0.922} & 0.837 \\
de\_de & \textbf{3.09} & 3.06 & \textbf{4.10} & 4.05 & \textbf{11.86} & 12.94 & \textbf{0.927} & 0.856 \\
el\_gr & \textbf{3.10} & 2.99 & \textbf{4.33} & 4.19 & \textbf{14.19} & 16.93 & \textbf{0.919} & 0.812 \\
en\_in & \textbf{3.02} & 2.97 & \textbf{4.31} & 4.19 & \textbf{33.46} & 35.23 & \textbf{0.932} & 0.841 \\
en\_us & \textbf{2.98} & 2.91 & \textbf{4.13} & 4.03 & \textbf{20.26} & 21.50 & \textbf{0.890} & 0.822 \\
es\_es & \textbf{3.22} & 3.13 & \textbf{4.10} & 4.05 & \textbf{8.28} & 9.10 & \textbf{0.938} & 0.831 \\
es\_us & \textbf{3.12} & 3.05 & \textbf{4.25} & 4.15 & \textbf{11.39} & 12.45 & \textbf{0.929} & 0.837 \\
et\_ee & \textbf{3.04} & 2.99 & \textbf{4.12} & 4.10 & \textbf{20.98} & 21.13 & \textbf{0.907} & 0.819 \\
fa\_ir & \textbf{3.18} & 3.09 & \textbf{4.09} & 3.88 & \textbf{13.28} & 16.25 & \textbf{0.925} & 0.847 \\
fi\_fi & \textbf{3.03} & 2.94 & \textbf{4.24} & 4.12 & \textbf{13.98} & 15.35 & \textbf{0.915} & 0.816 \\
fr\_ca & \textbf{3.12} & 3.05 & \textbf{4.21} & 4.12 & \textbf{18.20} & 19.36 & \textbf{0.928} & 0.844 \\
fr\_fr & \textbf{3.11} & 3.04 & \textbf{4.18} & 4.10 & 25.47 & \textbf{21.96} & \textbf{0.930} & 0.850 \\
gu\_in & \textbf{3.05} & 2.98 & \textbf{4.39} & 4.29 & \textbf{35.55} & 37.01 & \textbf{0.936} & 0.830 \\
hu\_hu & \textbf{3.08} & 2.89 & \textbf{4.18} & 4.05 & 40.32 & \textbf{36.09} & \textbf{0.907} & 0.835 \\
id\_id & \textbf{3.10} & 3.06 & \textbf{4.23} & 4.17 & \textbf{11.28} & 12.33 & \textbf{0.935} & 0.841 \\
it\_it & \textbf{3.15} & 3.11 & \textbf{4.04} & 4.02 & \textbf{11.14} & 12.23 & \textbf{0.936} & 0.837 \\
ja\_jp & \textbf{3.07} & 3.00 & \textbf{4.18} & 4.10 & \textbf{18.42} & 20.46 & \textbf{0.932} & 0.849 \\
km\_kh & \textbf{2.95} & 2.82 & \textbf{4.18} & 4.04 & 14.68 & \textbf{13.72} & \textbf{0.876} & 0.819 \\
ko\_kr & \textbf{3.08} & 3.03 & \textbf{4.40} & 4.26 & \textbf{23.55} & 27.66 & \textbf{0.925} & 0.822 \\
lt\_lt & \textbf{3.09} & 2.99 & \textbf{4.02} & 3.96 & \textbf{13.54} & 15.52 & \textbf{0.921} & 0.839 \\
lv\_lv & \textbf{3.25} & 3.21 & 4.05 & \textbf{4.06} & \textbf{7.33} & 7.84 & \textbf{0.941} & 0.850 \\
mr\_in & \textbf{3.09} & 2.98 & \textbf{4.42} & 4.32 & \textbf{28.25} & 30.25 & \textbf{0.926} & 0.839 \\
ms\_my & \textbf{3.15} & 3.07 & \textbf{4.12} & 4.11 & \textbf{14.09} & 15.64 & \textbf{0.945} & 0.866 \\
nb\_no & \textbf{2.91} & 2.86 & \textbf{4.33} & 4.19 & \textbf{18.35} & 21.60 & \textbf{0.920} & 0.806 \\
nl\_nl & \textbf{3.09} & 3.02 & \textbf{4.03} & 3.97 & \textbf{17.05} & 17.96 & \textbf{0.921} & 0.835 \\
pa\_in & \textbf{3.11} & 3.01 & \textbf{4.50} & 4.38 & \textbf{22.73} & 24.22 & \textbf{0.940} & 0.847 \\
pl\_pl & \textbf{3.07} & 3.03 & \textbf{4.31} & 4.23 & \textbf{10.71} & 12.46 & \textbf{0.925} & 0.833 \\
pt\_br & \textbf{3.05} & 2.94 & \textbf{4.14} & 4.05 & \textbf{10.34} & 12.26 & \textbf{0.923} & 0.813 \\
pt\_pt & \textbf{3.13} & 3.04 & \textbf{4.43} & 4.28 & \textbf{13.34} & 15.53 & \textbf{0.928} & 0.828 \\
ro\_ro & \textbf{3.20} & 2.94 & \textbf{4.11} & 4.00 & \textbf{7.30} & 9.15 & \textbf{0.910} & 0.821 \\
sk\_sk & \textbf{3.09} & 2.93 & \textbf{4.22} & 4.08 & \textbf{6.19} & 7.30 & \textbf{0.939} & 0.831 \\
sv\_se & \textbf{3.07} & 3.01 & \textbf{4.03} & 3.96 & \textbf{17.18} & 20.24 & \textbf{0.920} & 0.802 \\
ta\_in & \textbf{3.09} & 2.98 & \textbf{4.40} & 4.28 & \textbf{27.11} & 29.36 & \textbf{0.949} & 0.853 \\
th\_th & \textbf{3.07} & 2.98 & \textbf{4.32} & 4.18 & \textbf{12.09} & 13.19 & \textbf{0.933} & 0.820 \\
tr\_tr & \textbf{3.11} & 3.01 & \textbf{4.17} & 4.05 & \textbf{10.75} & 12.06 & \textbf{0.930} & 0.834 \\
uk\_ua & \textbf{3.07} & 3.01 & \textbf{4.17} & 4.07 & \textbf{10.82} & 12.18 & \textbf{0.928} & 0.838 \\
vi\_vn & \textbf{2.94} & 2.87 & \textbf{4.21} & 4.10 & 29.62 & \textbf{21.17} & \textbf{0.925} & 0.833 \\
\bottomrule
\end{tabular}
\end{table}

\section{Conclusion}

This paper introduced Miipher-2, a multilingual speech restoration model operating solely on noisy speech input without additional conditioning. The model surpasses a prior state-of-the-art monolingual English system in SR quality, measured by MOS, SxS, WER, and SPK, and computational efficiency, indicated by RTF and memory usage. Multilingual evaluations demonstrate its universal restoration capability in known/unknown languages. Finally, dataset distillation feasibility is shown, achieving nearly comparable performance by training Miipher-2 from scratch using only publicly available datasets.

The code and checkpoints will not be released due to potential misuse risks associated with recent advancements in generative models. Nevertheless, open-source reproduction would be feasible based on the methodology described herein, and by integrating public studio-quality multilingual datasets~\cite{fleursr,librittsr_koizumi23_interspeech} with pretrained multilingual speech encoders~\cite{boito2024mhubert,chen-etal-2024-towards-robust} and neural vocoders~\cite{nakata_miipher,ikemiya_wavefit}. %

\clearpage
\bibliographystyle{IEEEtran}
\bibliography{refs}

\end{document}